\documentstyle[preprint,aps]{revtex}

\begin{document} 
\draft

\title{Electromagnetic Energy for a Charged Kerr Black Hole \\
in a Uniform Magnetic Field}
\author{Li-Xin Li}
\address{Department of
Astrophysical Sciences, Princeton University, Princeton, NJ 08544}
\date{December 12, 1999}
\maketitle

\begin{abstract}
With the Komar mass formula we calculate the electromagnetic energy for a charged
Kerr black hole in a uniform magnetic field. We find that the total electromagnetic
energy takes the minimum when the Kerr black hole possesses a non-zero net charge
$Q = 2\xi B_0 J_H$ where $B_0$ is the strength of the magnetic field, $J_H$ is the
angular momentum of the black hole, $\xi$ is a dimensionless parameter determined
by the spin of the black hole.
\end{abstract}


Whether an astrophysical black hole can possess a net electric
charge is an interesting and important question. For an astrophysical black hole 
without magnetic field the answer seems to be clear: usually the black hole cannot 
possess much net electric charges since otherwise the black hole will 
selectively accrete particles with opposite sign of charges from ambient material and
be quickly neutralized \cite{mis73,wal84}. (However, if accretion onto a black hole
produces a luminosity close to Eddington limit, the black hole can acquire a 
net charge due to the different radiation drag on electrons and ions 
\cite{tre99}.) But for a black hole
immersed in a magnetic field --- which is believed to occur in many astrophysical
systems --- the situation is different and the answer is quite unclear. Wald 
has shown that when a Kerr black hole is immersed in a uniform magnetic
field aligned with its rotation axis, the hole acquires a net electric charge 
$Q_W = 2 B_0 J_H$, where $B_0$ is the strength of the magnetic field, $J_H$ is the 
angular momentum of the black hole (throughout the paper we use the geometric 
units $G = c = 1$) \cite{wal74}. Wald derived his
result from the requirement that the ``injection energy'' along the symmetry
axis should be zero for the equilibrium state. However, it can be shown that 
the ``injection energy'' defined by Wald in \cite{wal74} depends on the path along 
which the injection is made and the effect of off-axis accretion of charges is
unclear. So it remains a question what is the equilibrium state for a Kerr black hole
in a uniform magnetic field and whether the equilibrium state acquires a net charge.

Ruffini and Treves have analyzed the problem of a magnetized rotating sphere
in flat spacetime \cite{ruf73}. They have found that the sphere acquires a non-zero
net charge in order to minimize the total electromagnetic energy of the system.
This is similar to the conclusion of Wald but the acquired charge has opposite sign.
The method of extremizing the total energy is better than that of ``injection energy" 
since the former is a global approach which doesn't depend on the details of injection 
path and accretion process. Therefore in this paper we try to calculate the total
electromagnetic energy for a charged Kerr black hole in a uniform magnetic field.
Energy (mass) is well-defined for a stationary system which is asymptotically flat
and has an asymptotic timelike Killing
vector \cite{wal84}. However, if the uniform magnetic field extends to
infinity in space, the total electromagnetic energy diverges. To obtain a
finite electromagnetic energy we truncate the electromagnetic field with a
spherical surface. The black hole sits at the center of the sphere, the radius 
of the sphere is much larger than the radius of the black hole horizon. Inside the
sphere the magnetic field is uniform but outside the sphere the magnetic field
decreases quickly with increasing radius. (Currents are induced on the surface of the
sphere to connect the magnetic fields inside and outside the sphere.) We will 
calculate the total electromagnetic energy inside the sphere using the Komar mass
formula \cite{kom59} and show that when the total electromagnetic energy takes 
the minimum the black hole acquires a non-zero net charge $Q = 2\xi B_0 J_H = \xi Q_W$,
where the dimensionless parameter $\xi$ is a function of $a/M_H$ and
$0\leq\xi\leq\left[{3\over 2}\left(2+\pi\right)\right]^{-1}\approx 0.13$ for 
$0\leq a/M_H\leq1$, $M_H$ is the mass of the black hole and $a=J_H/M_H$.
(The value of $\xi$ is independent of the radius of the truncation sphere in the
limit that the radius of the sphere is much larger than the radius of the black hole.)
So the Wald state (which corresponds to $\xi=1$) doesn't minimize the total 
electromagnetic energy. Indeed, the energy of the Wald state is higher than that 
of the state with no charge.

For a charged rotating black hole immersed in a uniform magnetic field with
the electromagnetic field being sufficiently weak ($Q^2/M_H^2\ll 1$ and
$B_0^2 M_H^2\ll 1$), the spacetime can be described with Kerr metric. (In other
words, the electromagnetic field is treated as test field in the background of
Kerr spacetime.) In
Boyer-Lindquist coordinates, the Kerr metric is
\begin{eqnarray}
    ds^2 = -\left(1-{2M_Hr\over\Sigma}\right)dt^2-{4M_Har\over\Sigma}\sin^2\theta~dt
           d\phi+{\Sigma\over\Delta}dr^2+\Sigma d\theta^2+{C
           \sin^2\theta\over\Sigma} d\phi^2,
    \label{gab}
\end{eqnarray}
where $M_H$ is the Komar mass of the black hole, $a$ is the specific angular momentum
of the black hole (the angular momentum of the black hole is $J_H = M_H a$), and
\begin{eqnarray}
    \Delta = r^2-2M_H r+a^2, \hspace{1cm} \Sigma = r^2+a^2\cos^2\theta,
    \hspace{1cm} C = (r^2+a^2)^2-\Delta a^2 \sin^2\theta.
    \label{del}
\end{eqnarray}
The electromagnetic vector potential is \cite{wal74}
\begin{eqnarray}
    A^a = {B_0\over 2}\left[\left({\partial\over\partial\phi}\right)^a +
            2 a \left({\partial\over\partial t}\right)^a\right]
	    -{Q\over 2M_H}\left({\partial\over\partial t}\right)^a,
    \label{ap}
\end{eqnarray}
where $B_0$ is the strength of the external magnetic field and $Q$ is the
electric charge of the black hole. The electromagnetic field $F_{ab}$ is
given by
\begin{eqnarray}
    F_{ab} = \nabla_a A_b - \nabla_b A_a.
    \label{fab}
\end{eqnarray} 
The tensor indexes are raised and lowered with the Kerr metric $g_{ab}$.

Since $\left({\partial\over\partial t}\right)^a$ and $\left({\partial\over\partial
\phi}\right)^a$ are the Killing vectors of Kerr spacetime and the Ricci tensor
$R_{ab}=0$ for Kerr metric, $F_{ab}$ solves vacuum Maxwell's equations
$\nabla_a F^{ab} = 0$ \cite{wal74}. The stress-energy tensor of the electromagnetic
field is
\begin{eqnarray}
    T_{ab} = {1\over 4\pi}\left(F_{ac}F_b^{~c}-{1\over 4}g_{ab} F_{de}F^{de}\right).
    \label{tab}
\end{eqnarray}
The trace of the stress-energy tensor of the electromagnetic field is $T=
g_{ab}T^{ab} =0$.

With the Komar mass formula \cite{kom59}, the total mass (energy) for the Kerr
black hole and the electromagnetic field (with suitable truncation as described
earlier) is \cite{wal84,bar73,car79,fro98}
\begin{eqnarray}
    M = 2\int_{\Sigma}\left(T_{ab}-{1\over 2}Tg_{ab}\right)n^a \left({\partial
        \over\partial t}\right)^b dV +{1\over 4\pi}\kappa{\cal A}+
        2\Omega_H J_H,
    \label{kom}
\end{eqnarray}
where 
\begin{eqnarray}
    n^a = {1\over\alpha}\left[\left({\partial\over\partial t}\right)^a 
          +{2 a M_H r\over C}\left({\partial\over\partial\phi}\right)^a\right],
          \hspace{1cm} \alpha \equiv \left({\Delta\Sigma\over C}\right)^{1/2},
    \label{ua}
\end{eqnarray}
is the unit future-pointing normal to the hypersurface $\Sigma(t={\rm constant})$,
$dV$ is the volume element on $\Sigma$,
${\cal A}=4\pi\left(r_H^2+a^2\right)$ is the area of the horizon, $r_H=
M_H+\sqrt{M_H^2-a^2}$ is the radius of the horizon, $\kappa=\left(r_H-M_H\right)/
\left(2M_H r_H\right)$ is the surface gravity of the black hole, and $\Omega_H =
a/\left(2M_Hr_H\right)$ is the angular velocity of the black hole. So
the total energy of the electromagnetic field within the truncation sphere is
\begin{eqnarray}
    {\cal E}_{EM} = \int_{r_H}^R dr \int_0^\pi d\theta
                    \int_0^{2\pi} d\phi ~\epsilon\sqrt{h},
                    \hspace{1cm} \epsilon\equiv 2\left(T_{ab}-{1\over 2}Tg_{ab}
		    \right)n^a\left({\partial\over\partial t}\right)^b,
    \label{ene}
\end{eqnarray}
where 
\begin{eqnarray}
    \sqrt{h} = \left({\Sigma C\over \Delta}\right)^{1/2} \sin\theta
    \label{h}
\end{eqnarray}
is the measure of volume on $\Sigma$.
The integration over $r$ is truncated at $r = R$ (the radius of the truncation
sphere), otherwise the integration diverges since the magnetic field is 
asymptotically uniform. Suppose $R$ is $\gg r_H$ but $\ll \left(M_H/B_0^2
\right)^{1/3}$ [so that ${\cal E}_{EM}\ll M_H$ and the background spacetime can be
well approximated with the Kerr geometry]. The energy so defined is conserved 
since the spacetime is stationary.

Inserting Eqs.~(\ref{ap} - \ref{tab}) into Eq.~(\ref{ene}), we obtain
\begin{eqnarray}
    \epsilon = {\alpha\over 4\pi\Sigma^3}\left({f_0 B_0^2 + f_1 Q^2 +
               f_2 B_0Q}\right),
    \label{eps1}
\end{eqnarray}
where
\begin{eqnarray}
    f_0 &=&  r^5\left(r-2M_H\sin^2\theta\right)+ a^2 r^2\left[3r^2
             \cos^2\theta + 2M_Hr\left(1+\cos^2\theta\right)\left(
             1-3\cos^2\theta\right)\right. \cr
         &&  \left.-M_H^2\left(1+\cos^2\theta\right)\left(3-5\cos^2\theta\right)
             \right] +
             a^4\left[3r^2\cos^4\theta - 2M_Hr \cos^2\theta\left(3+\cos^2\theta
             \right)+\right.\cr
	 &&  \left.M_H^2\left(1+\cos^2\theta\right)^2\left(2-\cos^2\theta\right)
             \right]+a^6 \cos^6\theta,
    \label{f0}
\end{eqnarray}
\begin{eqnarray}
    f_1 = r^2+a^2\left(2-\cos^2\theta\right),
    \label{f1}
\end{eqnarray}
\begin{eqnarray}
    f_2 &=& -2a\left\{r^2\left(r-M_H\right)\left(1-3\cos^2\theta\right)-
            a^2\left[r\cos^2\theta\left(3-\cos^2\theta
            \right) - \right.\right.\cr
         && \left.\left.M_H\left(1+\cos^2\theta\right)\left(2-\cos^2\theta\right)
            \right]\right\},
    \label{f2}
\end{eqnarray}
and
\begin{eqnarray}
    {\cal E}_{EM} = {\cal E}_0 + \left({Q^2\over M_H}\right) F_1 +
    \left({B_0J_H Q\over M_H}\right) F_2,
    \label{ene1}
\end{eqnarray}
where (in the limit $R\gg r_H$)
\begin{eqnarray}
    {\cal E}_0 = {1\over 3} B_0^2 R^3 + {\cal O}\left(R^2\right),
    \label{ce0}
\end{eqnarray}
\begin{eqnarray}
    F_1 = F_1(s) = {1\over 2 s^3}\left(1-\sqrt{1-s^2}\right)
                   \left[s + 2\arctan\left({s\over {1+\sqrt{1-s^2}}}\right)\right]
                   + {\cal O}\left({1\over R}\right),
    \label{cf1}
\end{eqnarray}
\begin{eqnarray}
    F_2 = F_2(s) = {2\over 3 s^3}\left[s(3\sqrt{1-s^2}-s^2)-6(1-s^2)\arctan
                   \left({s\over {1+\sqrt{1-s^2}}}\right)\right]
                   + {\cal O}\left({1\over R^2}\right),
    \label{cf2}
\end{eqnarray}
where $s\equiv a/M_H$ is the spin parameter of the black hole. Since $0\leq s \leq 1$,
$F_1>0$ and $F_2\leq 0$ always. Since the total electromagnetic field is the
superposition of the external magnetic field $B_0$ and the electromagnetic field
generated by the charge $Q$, the total electromagnetic energy is composed of three
parts: (1) the ``bare" energy ${\cal E}_0$, i.e. the energy of the external magnetic 
field, which is proportional to $B_0^2$; (2) the energy of the electromagnetic field
generated by the charge $Q$, which is proportional to $Q^2$ (c.f. \cite{coh84} 
for the electromagnetic energy of a Kerr-Newman geometry); (3) the energy
arising from the interaction between the external magnetic field $B_0$ and the
electromagnetic field of the charge $Q$, which is proportional to
$B_0 Q$. Though as $R\rightarrow\infty$ the ``bare'' energy ${\cal E}_0$ diverges, 
$F_1$ and $F_2$ converge. So $F_1$ and $F_2$
don't depend on where the truncation is if $R\gg r_H$. In particular,
the difference in the electromagnetic energy between a charged Kerr black hole
in a uniform magnetic field and its uncharged state, $\Delta {\cal E} =
{\cal E}_{EM} -{\cal E}_0$, is independent of the truncation and so is well-defined.

${\cal E}_{EM}$ has a minimum since $F_1(s)>0$ for $0\le s\le 1$. By 
$\partial {\cal E}_{EM}/\partial Q = 0$ we obtain
\begin{eqnarray}
    Q = 2\xi(s)B_0 J_H,
    \label{nq}
\end{eqnarray}
where
\begin{eqnarray}
    \xi(s) = -{F_2\over 4F_1} = {s^3-3s\sqrt{1-s^2}+6(1-s^2)\arctan
              \left({s\over {1+\sqrt{1-s^2}}}\right)\over
              3\left(1-\sqrt{1-s^2}\right)\left[s + 2\arctan\left({s\over 
              {1+\sqrt{1-s^2}}}\right)\right]}.
    \label{xi}
\end{eqnarray}
For $0\leq s\leq 1$ we have $0\leq\xi\leq \left[{3\over 2}\left(2+
\pi\right)\right]^{-1}\approx 0.13$. $\xi(s)$ is plotted in Fig.~\ref{fig1}. 
$\xi$ decreases quickly as $s$ decreases. As examples: $\xi(0) = 0$, $\xi(0.1) 
\approx 3.4\times 10^{-4}$, $\xi(0.5) \approx 9.9\times 10^{-3}$, $\xi(0.9) 
\approx 5.6\times 10^{-2}$, $\xi(0.99) \approx 0.10$, and $\xi(1) \approx 0.13$.
So, a Kerr black hole immersed in a uniform magnetic field acquires a non-zero net
charge [given by Eq.~(\ref{nq})] to attain the minimum electromagnetic energy. 
Relative to the bare state (i.e. the state with $Q=0$), the value of the minimum 
electromagnetic energy is
\begin{eqnarray}
    \Delta{\cal E}_{min} = -\left({F_2^2\over 4 F_1}\right){B_0^2 J_H^2\over M_H}.
    \label{deq}
\end{eqnarray}
For a Kerr black hole with $a/M_H = 0.99$ we have $Q \approx 0.20 B_0 J_H$ and
$\Delta{\cal E}_{min} \approx -0.045 B_0^2 J_H^2/M_H$.

Wald's result $Q_W = 2 B_0 J_H$
does not correspond to the lowest energy state of the electromagnetic field.
In fact, the difference in the electromagnetic energy between the Wald state
and the uncharged state is
\begin{eqnarray}
   \Delta{\cal E}_{W} = 2\left(2F_1 + F_2\right){B_0^2 J_H^2\over M_H},
   \label{dew}
\end{eqnarray}
which is always positive for $0\leq a/M_H\leq 1$.
(For a Kerr black hole with $a/M_H = 0.99$ we have $\Delta{\cal E}_{W}
= 3.4 B_0^2 J_H^2/M_H$.) So the electromagnetic energy for the Wald state
is even higher than that of the uncharged state.

To see if the charge given by Eq.~(\ref{nq}) is important in practice, let's compare 
the contribution of the charge $Q$ and the magnetic field $B_0$ to the magnetic flux
through the northern hemi-sphere of the black hole horizon. The magnetic flux
contributed by $Q$ is $\Phi_Q = 2\pi a Q/r_H$. The magnetic flux contributed
by $B_0$ is $\Phi_{B_0} = 2\pi B_0 M_H r_H\left(1-a^2/r_H^2\right)$. (The total
magnetic flux is $\Phi = \Phi_Q+\Phi_{B_0}$.) For $Q$ given by Eq.~(\ref{nq}) we
have
\begin{eqnarray}
   {\Phi_Q\over\Phi_{B_0}} = {\xi s^2\over 1-s^2+\sqrt{1-s^2}},
   \label{rat}
\end{eqnarray}
which increases with increasing $s$. We find that $0\leq\Phi_Q/\Phi_{B_0} <1$ if
$0\leq s < 0.995$, $1< \Phi_Q/\Phi_{B_0} <\infty$ if $0.995<s\leq 1$. So
the charge given by Eq.~(\ref{nq}) is important only if $a/M>0.995$.

In conclusions, we have calculated the electromagnetic energy for a charged Kerr
black hole immersed in a uniform magnetic field. We have found that a non-zero
net charge is acquired by the Kerr black hole for attaining the minimum electromagnetic
energy. The Wald state isn't the state with minimum electromagnetic energy. Indeed
the electromagnetic energy for the Wald state is even higher than that for the
uncharged state. Though the realistic case for an astrophysical black hole is
much more complicated than the simple model investigated here due to the
appearance of many charged particles in the neighborhood of the black hole,
our results have shown that a Kerr black hole in the lowest energy state in
an external magnetic field acquires a non-zero net charge. 

\acknowledgments{I am very grateful to Robert M. Wald for many helpful and
stimulating discussions. This work was supported by the NSF grant 
AST-9819787.}

\begin{figure}
\caption{A Kerr black hole immersed in a uniform magnetic field acquires a net 
charge $Q = 2\xi B_0 J_H$ when the total electromagnetic energy takes the minimum. 
The dimensionless parameter $\xi$ is a function of $s=a/M_H$, which is shown with
the solid curve. For reference the Wald charge ($\xi=1$) is also shown with the
dashed line, which doesn't correspond to the minimum energy of the electromagnetic
field.}
\label{fig1}
\end{figure}

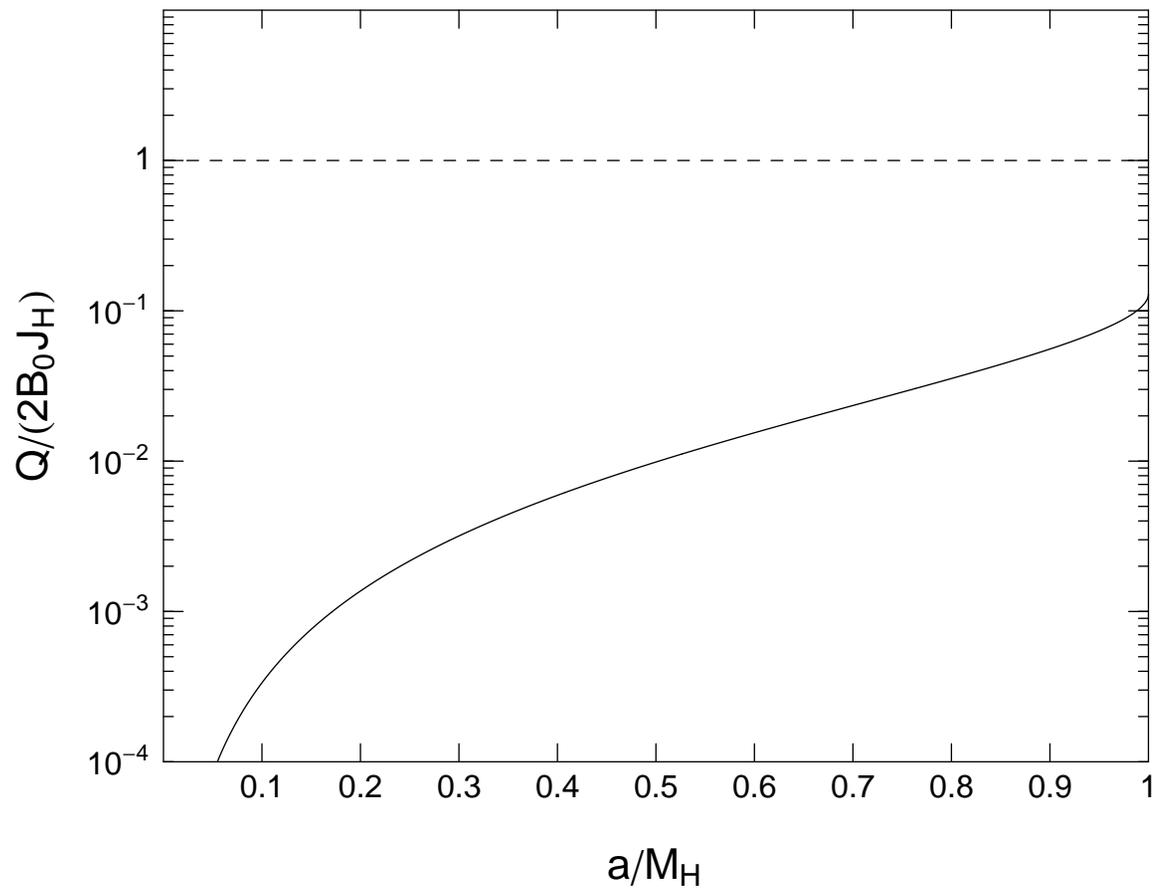
\end{document}